\documentclass[prl,twocolumn,aps,superscriptaddress, longbibliography]{revtex4-2}
\usepackage{amsmath}
\usepackage{amssymb}
\usepackage{todonotes}
\usepackage[unicode=true,colorlinks=true,citecolor=blue,urlcolor=blue]{hyperref}
\usepackage{float}
\usepackage{bm}
\usepackage{setspace}
\usepackage{epsfig}
\usepackage{lipsum}
\usepackage{physics}
\usepackage{comment}
\renewcommand {\phi}{{\varphi}}
\newcommand {\rmi}{{\rm i}}

\newcommand {\e}{{\rm e}}

\renewcommand{\addcontentsline}[3]{}%

\begin{document}

\title{No oscillating subradiant correlations in  a strongly driven quantum emitter array
}

\author{Jiaming Shi}
\affiliation{Department of Physics of Complex Systems, Weizmann Institute of Science, Rehovot 7610001, Israel}
\email{jiaming.shi@weizmann.ac.il}

\author{Nikita Leppenen}
\affiliation{Department of Chemical \& Biological Physics, Weizmann Institute of Science, Rehovot 7610001, Israel}

\author{Ran Tessler}
\affiliation{Department of Mathematics, Weizmann Institute of Science, Rehovot 7610001, Israel}
\email{ran.tessler@weizmann.ac.il}

\author{Alexander N. Poddubny}
\email{poddubny@weizmann.ac.il}
\affiliation{Department of Physics of Complex Systems, Weizmann Institute of Science, Rehovot 7610001, Israel}

\begin{abstract}
We theoretically study time-dependent correlations in a strongly driven array of $N$ two-level atoms, coupled to photons in a waveguide.  We focus on the spectrum $\{\lambda\}$ of the Liouvillian superoperator, which determines the correlation decay rates $-\Re \lambda$ and the frequencies $\Im\lambda$.  Our main finding is {the} suppression of  subradiant oscillating correlations  between atomic states by  a strong coherent drive {of} amplitude $\Omega$:
$|\Re \lambda|\ge m\gamma/2$, where  $\gamma$ is {the} single-atom spontaneous decay rate and 
{$m=|\Im \lambda/(2\Omega)|$} is a nonzero integer for correlations oscillating in time $\propto \exp(\pm 2\rmi m|\Omega| t)$. This limit is independent of the number of atoms $N$; it holds both for small arrays and in the macroscopic limit. 
We demonstrate the suppression of subradiance numerically and provide a rigorous proof based on the {analytical decomposition }of the Liouvillian using spectral theory of simplicial complexes and posets.
\end{abstract}

\date{\today}
\maketitle

\begin{figure}[b]
\centering
\includegraphics[width=\linewidth]{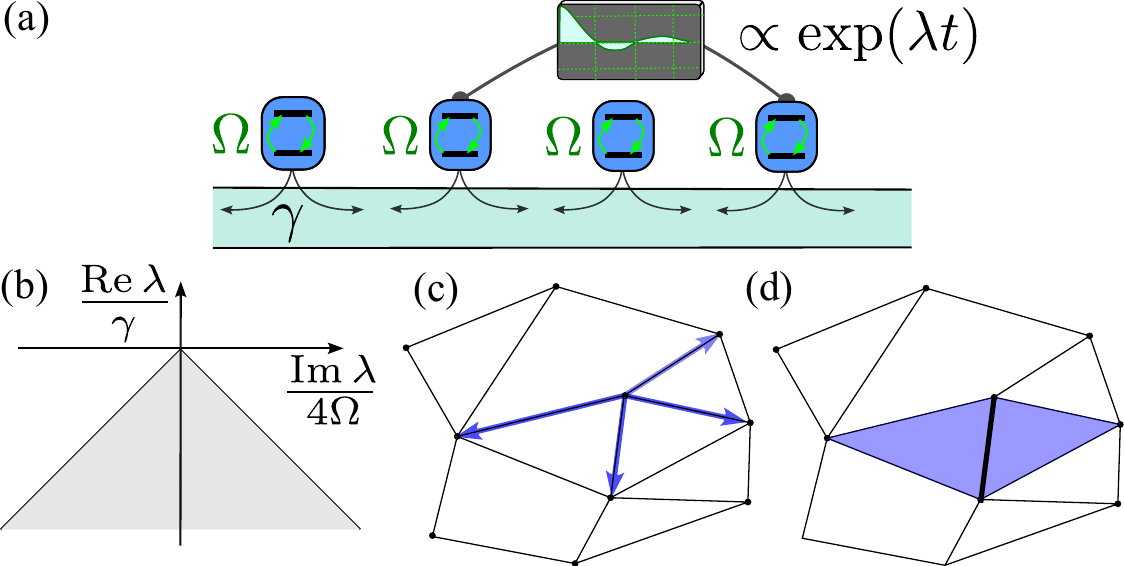}
\caption{(a) Schematics of the array of atoms coherently driven with  Rabi frequency $\Omega$ and dissipating photons into the waveguide. (b) Spectral bound Eq.~\eqref{eq:limit} on the complex plane of Liouvillian eigenvalues $\lambda$. (c,d)  Illustration of the  oriented incidence matrix  $\mathcal B$ for the graph Laplacian (the matrix maps vertices to the edges)  (c) and for the Laplacian of a simplicial complex (the matrix maps edges to adjacent faces) (d)
}
\label{fig:1}
\end{figure}

An ensemble of quantum emitters,  driven by an electric field and dissipating energy by radiating photons, is one of the paradigmatic systems to study driven-dissipative dynamics and many-body physics. Its temporal dynamics can be described by the Lindblad equation for the emitters' density matrix $\rho$, 
$\partial_t\rho=\mathcal L\rho$, where the $\mathcal L$ is the Liouvillian superoperator~\cite{Lehmberg}. The spectrum $\{\lambda\}$ of the Liouvillian characterizes the decay rates and oscillation frequencies of the correlations.  The long-living correlations with small $|\Re\lambda|$ describe long-time relaxation to the stationary state and are therefore of special significance. Depending on the system parameters, the spectrum can have various special points~\cite{Albert_2016,Minganti_2019}. 
Exceptional points, where the eigenvectors of the Liouvillian coalesce, correspond to the dissipative phase transitions~\cite{Werner2005,Capriotti2005,Kessler2012,Lee2014,Barberena2019}. Time-dependent oscillating solutions, with $\Im\lambda\ne 0$ and $\Re \lambda\to 0$ in the macroscopic limit, describe the so-called time-crystalline phase~\cite{Iemini2018}.
{The general structure of the Liovillian spectra still remains an open theoretical problem under active development~ \cite{Baumgartner_2008,Nigro_2019,McDonald_2023}.}

\begin{figure}[t]
\centering
\includegraphics[width=\linewidth]{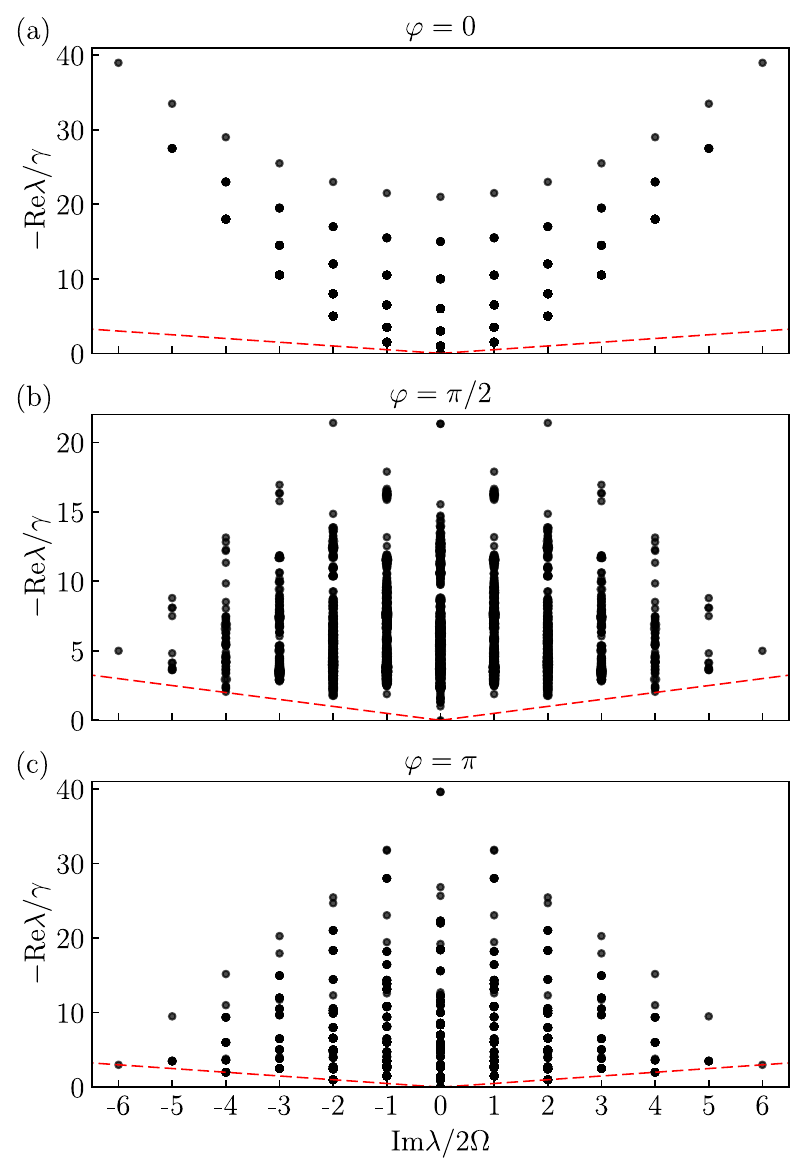}
\caption{Liouvillian spectrum of a 6-atom {equidistant} array for an array period  $\phi=0$ (a),  $\phi=\pi/2$ (b), and  $\phi=\pi$ (c). The red dashed lines show the spectral bound {of} Eq.~\eqref{eq:limit}. The calculation has been performed for $\Omega=100$, and $\gamma  =1$. }
\label{fig:spectrum}
\end{figure}

In this Letter, we consider the oscillating part of the spectrum, $\Im \lambda\ne 0$, for the array where identical atoms
are coupled to the waveguide and 
are coherently resonantly driven by the field with the same amplitude (characterized by the  Rabi frequency $\Omega$), see Fig.~\ref{fig:1}(a). In the strong driving limit, 
the eigenfrequencies $\Im \lambda$ are {approximately} equal to the integer multiples of $2\Omega$. The typical Liouvillian spectrum is shown in Fig.~\ref{fig:spectrum}, {where} the eigenvalues are separated into groups aligned along the vertical lines $\Im \lambda=0,\pm2\Omega,$ etc. Due to the photon emission, all the eigenstates except the  single stationary state have  nonzero decay rates 
$-\Re \lambda$. Because of the constructive or destructive interference of photons from different emitters, the decay rates can, in principle, be both larger or smaller than the individual decay rate of a single atom $\gamma$ --- the latter case would correspond to subradiance.
Subradiant eigenstates of non-Hermitian Hamiltonians have been studied in detail in the recent years, both theoretically \cite{Ostermann_2019,Molmer2019,poddubny2020quasiflat,poshakinskiy_dimerization_2021,Molmer2022,Volkov2023,Shi2024} and experimentally
\cite{2019arXiv190713468W,brehm2020waveguide,zanner2021coherent,Ferioli_2021,Glicenstein_2022,Tiranov2023,Kim_2024}. Their counterparts, subradiant eigenstates of the Liouvillians for driven atom-photon coupled systems with $|\Re \lambda|\ll \gamma$, are much less understood.

Our main result, illustrated  in Fig.~\ref{fig:1}(b) and shown by the calculations in Fig.~\ref{fig:spectrum}, is the {\it absence of oscillating subradiant correlations} for a strong driving:
\begin{equation}\label{eq:limit}
    |\Re \lambda| {\geq} \frac{\gamma m}{2},\quad  m=\frac{|\Im \lambda|}{2|\Omega|}\:.
\end{equation}
For strong driving, $|\Omega|\gg N\gamma$ {(where the factor $N$ takes into account possible collective enhancement of emission)}, $m$ is an integer. If  $m$ is nonzero i.e., correlations oscillate in time $\propto \e^{\pm2\rmi m|\Omega| t}$, then the decay rate $|\Re \lambda|$ is bounded from below. For a given $m$, this bound does not depend on the number of atoms $N$.   For the non-oscillating part of the spectrum with $\Im\lim_{\Omega\to\infty} \lambda/|\Omega|=0$, considered so far~\cite{poddubny_2022_driven,leppenen2025persistentsubradiantcorrelationsrandom,Robicheau2024}, including our own works \cite{poddubny_2022_driven,leppenen2025persistentsubradiantcorrelationsrandom}, Eq.~\eqref{eq:limit} does not set any restrictions, since $m=0$.
The points at the central line $\Im\lambda=0$ can be below $\gamma$ for large {$|\Omega|$},  see Fig.~\ref{fig:spectrum}.
That is, non-oscillating correlations can be subradiant, only oscillating ones can not.
The proposed limit has also not been discussed for the so-called Dicke boundary time crystal phase. There, the  solutions with $\Re \lambda\to 0$ appear only because the decay rate $\gamma$ is rescaled as $\gamma\to\gamma/\sqrt{N}$ in the macroscopic limit of an array in a cavity~\cite{Iemini2018} and without such a rescaling, the limit Eq.~\eqref{eq:limit} still holds.
{The limits for $|\Re \lambda|$ for nonzero $\Im \lambda$ are  known in classical statistical physics~
\cite{Barato_2017,Ohga_2023,xu2025thermodynamicgeometricconstraintspectrum}:  Eq.~\eqref{eq:limit} has the same form as the bound by Barato and Seifert~\cite{Barato_2017}, proved in Ref.~\cite{Ohga_2023}. Here, however, we consider not a classical model characterized by rate equations but a fully quantum setup, that is coherently driven and has collective dissipation to the photonic bath with zero temperature.  Therefore, while our limit has the same general form as in  Ref.~\cite{Ohga_2023}, our quantum model is very different and requires a completely independent consideration. }

Suppression of subradiant correlations by a strong driving may still have a simple physical interpretation already for an $N=1$ atom with a ground state $|g\rangle$ and an excited state $|e\rangle$.
The ground state has an infinite lifetime, while the excited state can spontaneously decay to the ground state with the rate $\gamma$. The strong driving leads to the Rabi oscillations between these two states, or, in other words, it leads to the new hybridized eigenstates $(|e\rangle \pm|g\rangle)/\sqrt{2}$ in the rotating frame. Since both hybridized states have a nonzero contribution from the excited state, their lifetime has to be finite. This argument naturally extends to the multiple atoms in the Dicke model, where the driving couples collective eigenstates with the same total momentum  $J$ and different excitation number $(N/2)+J_z$~\cite{Damanet2016,shammah_superradiance_2017,leppenen2024}. For a given $J$ only the eigenstate with the lowest momentum projection $J_z=-J$  is subradiant, and the rest have a finite lifetime. Based on  this argument  one can not expect the correlations with infinite lifetime. Ideal destructive interference appears to be incompatible with the Rabi oscillations.  This argument still does not prove an existence of any rigorous bound and it is also unclear why it should hold beyond the Dicke model when the spherical symmetry is broken. 

In this Letter, we rigorously  prove the existence of the spectral limit  {of} Eq.~\eqref{eq:limit} for the waveguide quantum electrodynamics (QED) setup: an array of atoms, coupled to propagating photons in a waveguide~\cite{poddubny2023RevModPhys}. 
Our findings are corroborated by the numerical calculations of the Liouvillian spectrum in a wide range of parameters. 
The very possibility of an exact proof in the strong driving limit is far from obvious, since even without the driving, this model is not integrable and shows chaotic eigenstates~\cite{Poshakinskiy2021chaos}.
Our analytical proof relies on the decomposition of  the Liouvillian terms, describing driving and dissipation,  in the subspace with fixed $\Im\lambda$  as $\mathcal L=- \mathcal F-\mathcal A$, where   $\mathcal F$  is a positive definite diagonal operator in the basis of the eigenstates of the driving Hamiltonian $H_{\rm I}$, and 
$\mathcal A$  is also positive semi-definite.  
Notably, the  operator { $\mathcal{A}$} has the form of a scaled Laplacian of a ranked poset~\cite{kalai1983enumeration,horak2011spectracombinatoriallaplaceoperators,mulas2022graphs,steenbergen2013towards,oppenheim2018local,kaufman_et_al:LIPIcs.ITCS.2023.78} constructed on the eigenstates of the driving Hamiltonian, $\mathcal A=\mathcal B^T\mathcal B$.

{\it Simplicial complexes and ranked posets}. To provide more background, a \emph{partially ordered set (poset)} is a set $\mathcal{P}=\{p_1,\ldots,p_m\},$ together with a partial order $<$ which satisfies that if $p_i<p_j$ and $p_j<p_k$ then $p_i<p_k,$ and never $p_i<p_i.$
We say that $p_j$ \emph{covers} $p_i$ if $p_i<p_j$ and there is no intermediate $p_k$ with $p_i<p_k<p_j$. A poset is \emph{ranked} if there is a rank function $r:\mathcal{P}\to\{0,1,2,\ldots\}$ such that $r(p_j)=r(p_i)+1$ whenever $p_j$ covers $p_i.$
We write $\mathcal{P}_k$ for the subset of $\mathcal{P}$  made of elements of rank $r=k.$

Prominent examples of ranked posets are provided by simplicial complexes. A simplicial complex on a base set $X$ is a collection of subsets $\mathbf{S}=\{S_1,\ldots,S_m\subseteq X\}$ called \emph{simplices}, which is \emph{closed to taking subsets}, that is, if $S\in \mathbf{S}$ then also every subset of $S$ belongs to $\mathbf{S}.$ 
The \emph{rank} of $S\in\mathbf{S}$ is defined to be $|S|-1,$ the cardinality of $S$ minus $1$.
Simplicial complexes admit a ranked poset structure where $<$ is given by set containment $\subset$, and $r$ is the simplex rank.

A \emph{rank $k$ weighted incidence operator} is an operator $B_k: (\mathbb{R}^a)^{\mathcal{P}_k}\to (\mathbb{R}^b)^{\mathcal{P}_{k+1}},$ where by $(\mathbb{R}^a)^{\mathcal{P}_k}$ we mean the vector space $\mathbb{R}^{a|\mathcal{P}_k|}$, whose coordinates are labeled $(c,p)$ for $c\in\{1,\ldots, a\},~p\in\mathcal{P}_k,$ and we similarly define $(\mathbb{R}^b)^{\mathcal{P}_{k+1}}.$ We require all entries of the form $[(c,p')(c',p)]$ to be zero, unless $p'< p.$ For example, in Fig.~\ref{fig:1}(c,d) we illustrate a simplicial complex, together with a vertex with adjacent edges ($k=1$), and an edge with adjacent faces ($k=2$).
Generally, a \emph{rank $k$ weighted Laplacian} is an operator of the form $\mathcal B^T\mathcal B:(\mathbb{R}^a)^{\mathcal{P}_k}\to(\mathbb{R}^a)^{\mathcal{P}_k}$ where $\mathcal B$ is a rank $k$ weighted incidence matrix. 

In our case, the operator  $\mathcal B$ connects ranked posets built on the degenerate sets of the eigenstates of $H_{\rm I}$ with the eigenvalues differing by {$2|\Omega|$}. The study of spectral theory of weighted Laplacians in simplicial complexes and ranked posets has attracted considerable attention in recent years \cite{kalai1983enumeration,horak2011spectracombinatoriallaplaceoperators,mulas2022graphs,steenbergen2013towards,oppenheim2018local,dikstein2024boolean,kaufman_et_al:LIPIcs.ITCS.2023.78}. It has found applications in the study of high dimensional expanders in mathematics \cite{lubotzky2005ramanujan,lubotzky2005explicit,parzanchevski2016isoperimetric,kaufman2016isoperimetric,evra2024bounded}, in complexity theory \cite{khot2017independent,dinur2018non,dinur2018towards,subhash2018pseudorandom}, error correcting codes (classical and quantum) \cite{dinur2022locally,panteleev2022asymptotically,leverrier2023decoding} and efficient sampling \cite{anari2018log,anari2019log,alev2020improved,anari2021entropic,alimohammadi2021fractionally} in theoretical computer science. Here, we apply this decomposition to quantum optics: by knowing from numerical calculations that $\mathcal A$ is positively defined, we were specifically looking for the decomposition $\mathcal A=\mathcal B^T\mathcal B$.

{\it Model.}
The Liouvillian superoperator, governing the dynamics of the atomic array coupled to photons, can be presented as~\cite{Blais2013,Caneva_2015,pichler2015quantum}: 
\begin{equation}\label{eq:master}
    \mathcal{L}\rho =  -\rmi[H_{\rm I} + H_{\rm II}, \rho] + \gamma \mathcal{D}[c_L]\rho + \gamma \mathcal{D}[c_R]\rho \:,
\end{equation}
where $[a,b]=ab-ba$ is the commutator, the dissipators are {$\mathcal{D}[L]\rho\equiv L\rho L^\dagger - (L^\dagger L\rho+\rho L^\dagger L)/2$,} and the Hamiltonian is
\begin{align}\label{eq:H}
    H_{\rm I} &= - \sum_{j=1}^{N}(  \Omega \sigma_j^\dagger + \Omega^* \sigma_j^{\vphantom{\dagger}} ) \nonumber \:, \\
    H_{\rm I\!I} &= -\frac{\rmi  \gamma}{2} \sum_{j\neq l} (\e^{\rmi k|z_j-z_l|}\sigma_j^\dagger \sigma_l^{\vphantom{\dagger}} - \text{H.c.})\:.
\end{align}
Here, we assume $\hbar=1$,  $\sigma_j,\sigma_j^\dag$ are the usual lowering and raising operators, operating in the basis of ground $|g_j\rangle$ states and 
excited $|e_j\rangle$ states of $j$-th atom, $\sigma_j|e_j\rangle=|g_j\rangle$, $k$ is the wavevector of photon in waveguide and  $z_j$ is the position of $j$-th atom. The collective jump operators describing spontaneous emission to the left and to the right are $c_L = \sum_{j=1}^N \e^{\rmi \phi_j}\sigma_j$, $c_R = \sum_{j=1}^N \e^{-\rmi \phi_j}\sigma_j$, and $\varphi_j=k z_j$. The Hamiltonian  $H_{\rm I}$ describes the coherent driving characterized by the Rabi frequency $\Omega$, and the  $H_{\rm I\!I}$ describes waveguide-mediated coupling between the atoms. 
We assume that the driving is resonant with the emitter frequency and use the rotating wave approximation and the rotating frame, so the Liouvillian Eq.~\eqref{eq:master} is time-independent~\cite{Iemini2018,leppenen2024}. 
The Liouvillian Eq.~\eqref{eq:master} also relies on the   Markovian approximation. 

We are interested in the Liouvillian spectrum $\mathcal L\rho=\lambda\rho$ in the limit of strong driving, $\Omega\gg {N}\gamma$. 
The results of full numerical realization for several values of array period $\varphi_n-\varphi_{n-1}=\varphi$ are shown in {Fig.~\ref{fig:spectrum}(a--c).
As expected, for large $\Omega\gg \gamma$, the spectrum splits into distinct branches with integer
$m=\Im\lambda/(2|\Omega|)=-N,-N+1\ldots N$.  This is because the eigenstates of Liouvillian are of the type $|w\rangle\langle w'|$, where $|w\rangle,|w'\rangle$ are the eigenstates of the operator $H_{\rm I}=-2\Omega J_x$, that is proportional to the total angular momentum projection operator $J_x=\sum_{j=1}^N(\sigma_j^{\vphantom{\dag}}+\sigma_j^\dag)/2$ with the discrete eigenvalues $J_x=-N/2,-N/2+1\ldots N/2$.
The results in Fig.~\ref{fig:spectrum} clearly indicate that the spectrum of the Liouvillian lies above the limit of Eq.~\eqref{eq:limit}, {which is indicated} by the red dashed lines. We have considered, for simplicity, a periodic array,
and we set the same driving field phase at all the atoms.  However,   Eq.~\eqref{eq:limit} applies for arbitrary values of $\varphi_n$ and for arbitrary driving phases
We also present in Fig.~\ref{fig:scan} the dependence of the minimum decay rates $|\Re \lambda|$ on the array period $\varphi$ for several spectral branches with distinct $m$  and $N=4\ldots 6$. The calculation shows that the minimal values of $|\Re \lambda|$ are always larger than $m\gamma/2$, satisfying Eq.~\eqref{eq:limit}.
Solid and dashed curves are calculated for the full Liouvillian and the Liouvillian where the interaction term 
$\propto H_{\rm I\! I}$ has been neglected. 
The solid curves lie above unless $\varphi=0,\pi,2\pi$, when  $ H_{\rm I\! I}\equiv 0$. This indicates that the interaction 
term $\propto H_{\rm I\! I}$ increases $|\Re \lambda|$, making the  correlations decay faster.

{\it Proof of the limit Eq.~\eqref{eq:limit}.} We start the proof by projecting the Liouvillian to the subspaces of the eigenstates of the driving Hamiltonian $H_{\rm I}$. 
We first define the states with effective spin parallel and antiparallel to the $x$ direction, $|1\rangle = \frac{1}{\sqrt{2}} (|g\rangle + |e\rangle)$ and $|0\rangle = \frac{1}{\sqrt{2}}(|g\rangle - |e\rangle)$. We will use a binary word $w$ (e.g. 011) to describe a direct product state $|w\rangle$ (e.g. $|011\rangle = |0\rangle \otimes |1\rangle \otimes|1\rangle$), which is an eigenstate of the driving Hamiltonian $H_{\rm I}$. We define the value of a binary word by summing all the bits $s_w = \sum_{j=1}^N w_j$, where $w_j$ is the value of the $j$-th index. Furthermore, we use a binary word pair $(w,w')$ to describe a density matrix $\rho = |w\rangle\langle w'|$. Obviously, $-\rmi[H_{\rm I}, \rho] = 2\rmi \Omega  (s_w - s_{w'})\rho$. 
In the limit of strong driving, the eigenspace of the full Liouvillian is split into subspaces $\{\rho_i\}\equiv \{ |w\rangle \langle w'|\}$ with different integer values of $s_w - s_{w'} = m$. {Without the loss of generality, we assume $m\geq0$.} In this reduced basis  $\{\rho_i\}$ the dissipators are represented  by a  matrix $\mathcal{Q}_{ij} = \Tr\{\rho_i (\mathcal{D}[c_L] +  \mathcal{D}[c_R])\rho_j\}$. 

\begin{figure}[t]
\centering
\includegraphics[width=\linewidth]{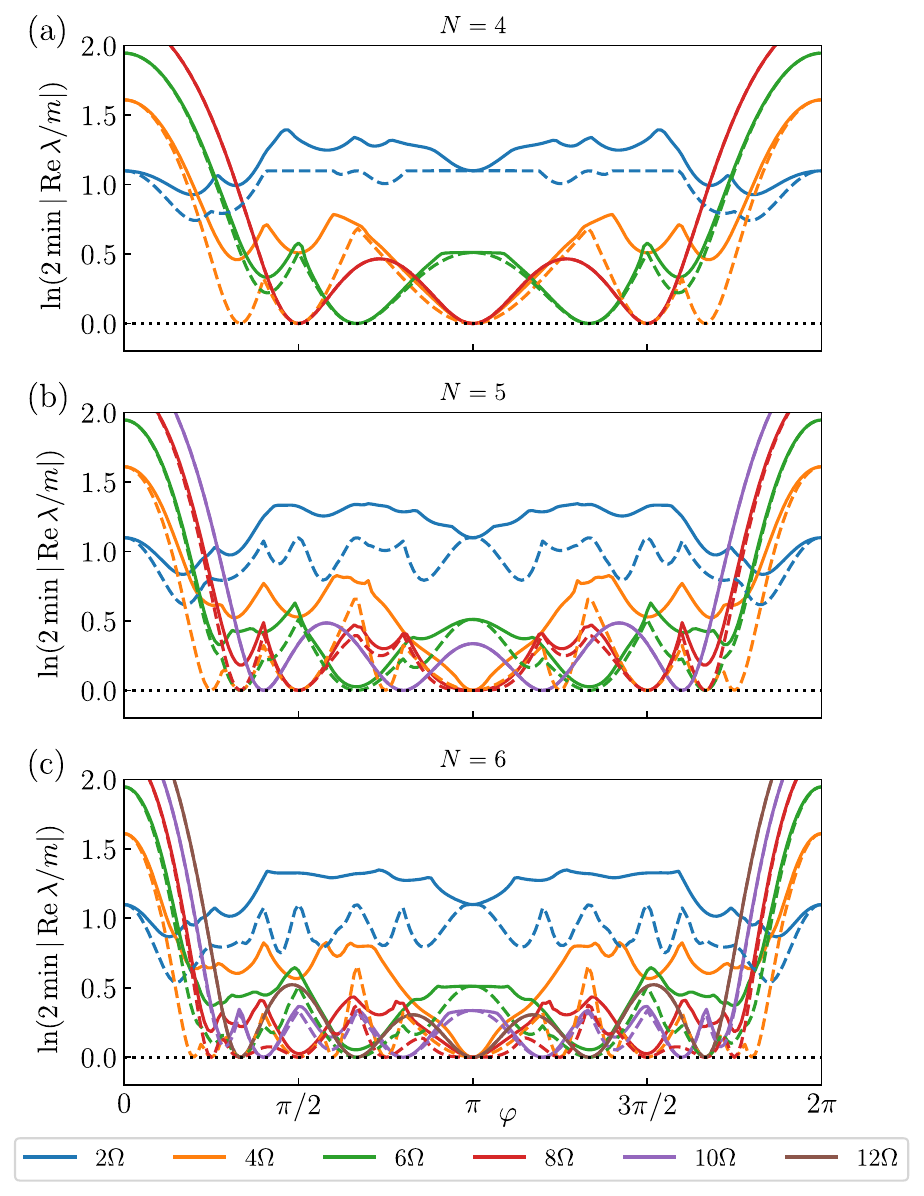}
\caption{Dependence of the minimal absolute value of the real part of the Liouvillian eigenvalues  on the array period $\varphi$ for (a) $N=4$, (b) $N=5$, and (c) $N=6$.
Different curves correspond to the different values of $\Im\lambda$, indicated on the plot.
Dashed lines represent the results without the interaction term $H_{\rm I\!I}$.  The calculation has been performed for $\Omega=100$, and $\gamma  =1$. }
\label{fig:scan}
\end{figure}

We now decompose $\mathcal{Q}$ as a sum of a diagonal matrix $\mathcal{F}$ and a symmetric matrix $\mathcal{A} $, $\mathcal{Q} = -\mathcal{F}-\mathcal{A}$. The off-diagonal term of $\mathcal{A}$ is
\begin{equation}\label{eq:matrixA}
    \begin{split}
        \mathcal{A}_{\rho, \rho'} =
        \frac{1}{2}\sum_{i,j}\cos(\phi_i-\phi_j)
        (- \chi_{j,j}+\delta_{w',v'}\xi_{i,j} +
        \delta_{w,v}\xi'_{i,j}) \:,
    \end{split}
\end{equation}
where $\rho=|w\rangle \langle w'|, \rho' = |v\rangle\langle v'|$. Here, we introduced the indicator function  $\chi_{i,j}$ that is equal to unity if and only if $w,v$ differ at exactly one index $i$ while $w',v'$ differ at exactly one index $j$; otherwise $\chi_{i,j}=0$. Similarly, $\xi_{i,j} = 1$ if and only if $w$ and $v$ differ at exactly indices $i$ and $j$; otherwise $\xi_{i,j}=0$ and $\xi'_{i,j} = 1$ if and only if $w',v'$ differ at exactly indices $i$ and $j$; otherwise $\xi'_{i,j}=0$.

Our proof relies on the observation that if the diagonal term of $\mathcal{A}$ is set to $(N-m)/2$, then $\mathcal{A}$ becomes a weighted Laplacian of a ranked poset $\mathcal{P}$, hence is positive semi definite. The poset of interest consists of elements 
{$(w,w')$} with $s_w\geq s_{w'}.$ This set becomes a ranked poset by defining the rank function  $r(w,w')=s_w-s_{w'}$, and declaring 
$(w,w')<(t,t')$ if $r(w,w')<r(t,t'),$ {and} for every index $i$ with $w_i=1$ {there is} also $t_i=1,$ and for every index $i$ with $t'_i=1$ {there is} also $w'_i=1.$  
The operator $\mathcal{A}$ can then be decomposed as $\mathcal A=\mathcal B^T\mathcal B$, where $\mathcal{B}$ is the weighted incidence operator described as follows. When $m=N$, ${|w\rangle\langle w'|}$ has only one element $|11\dots 1\rangle\langle 00\cdots0|$ and $\mathcal{A}=\mathcal{B}=0$. When $0\leq m<N$, the elements of $\mathcal{B}:\mathbb{R}^{|\mathcal{P}_m|}\to\mathbb{R}^{2|\mathcal{P}_{m+1}|}$ are determined by the 
{posets $\{(w,w')\}$ with $r(w,w') = m$ and $\{(t,t')\}$ with $r(t,t') = m+1$.} {
The physical meaning of   $\mathcal B$  is thus an effective raising operator, changing collective momentum projection $J_x$ along the drive direction by unity. } If {$(w,w')_j <(t,t')_i$ and they} differ at exactly one index $k$, then $\mathcal{B}_{ij} = \eta \boldsymbol{\theta}_k$, where $\eta=1$ if $k$ is in the left word and $\eta = -1$ if $k$ is in the right word and the vector $\boldsymbol{\theta}_k \equiv (\cos\phi_k, \sin\phi_k)^T/\sqrt{2}$. Otherwise $\mathcal{B}_{ij} = (0,0)^T \equiv \mathbf{0}$. An explicit proof of this decomposition is given in the End Matter.

For instance, when $N=2$ and $m=0$, $\{(w, w')\}$ contains (11,11), (10,10), (10,01), (01,10), (01,01), (00,00) and $\{(t,t')\}$ contains (11,10), (11,01), (10,00), (01,00). The matrix $\mathcal{B}$ is
\begin{equation}
\mathcal{B} =
\begin{pmatrix}
-\boldsymbol{\theta}_{2} &  \boldsymbol{\theta}_{2} & \mathbf{0} & \boldsymbol{\theta}_{1} & \mathbf{0} & \mathbf{0} \\
-\boldsymbol{\theta}_{1} &  \mathbf{0} & \boldsymbol{\theta}_{2} & \mathbf{0} & \boldsymbol{\theta}_{1} & \mathbf{0} \\
\mathbf{0}  & -\boldsymbol{\theta}_{1} & -\boldsymbol{\theta}_{2} & \mathbf{0} & \mathbf{0} & \boldsymbol{\theta}_{1} \\
\mathbf{0}  &  \mathbf{0} & \mathbf{0} & -\boldsymbol{\theta}_{1} & -\boldsymbol{\theta}_{2} & \boldsymbol{\theta}_{2}
\end{pmatrix}\:.
\end{equation}
The matrix element $\mathcal B_{21}$ is determined by the word pairs (11,11) and (11,01). The difference is in the first index of the right word. Therefore $\mathcal B_{21} = -\boldsymbol{\theta}_1$.

The diagonal matrix $\mathcal{F}$ is given by
\begin{equation}
    \mathcal{F}_{\rho} = \frac{1}{4}|a_w-a_{w'}|^2 + \frac{m}{2}\:,
\end{equation}
where $a_w = \sum_{j=1}^N (-1)^{w_j} \e^{\rmi \phi_j}$.

The obtained decomposition $\mathcal{Q} = -\mathcal{F}-\mathcal{B}^T\mathcal B$  proves that in the reduced basis of density matrices {with $r(w,w') = m$}, corresponding to $|\Im \lambda|=2m\Omega$, the eigenvalues of $\mathcal{D}[c_L] + \mathcal D[c_R]$ are not larger than $-{m}/{2}$. We notice that the superoperator $\mathcal{D}[c_L] + \mathcal D[c_R]$ is Hermitian while $-\rmi [H_{\rm I\!I}, \cdot \,]$ is  anti-Hermitian. Therefore, using the Bendixson inequality~\cite{Bendixson1902}, we find that the real parts of eigenvalues $\lambda$ for the total Liouvillian, given by a sum of these two operators, should be no larger than $-\frac{m}{2}$. This concludes the proof of our limit Eq.~\eqref{eq:limit}. It also
 explains why the solid curves in Fig.~\ref{fig:scan}, calculated including the interaction term $H_{\rm I\! I}$, are always above the corresponding dashed curves, calculated neglecting this term.  Since the proof
 does not rely on the specific form of  $H_{\rm I\!I}$, the limit Eq.~\eqref{eq:limit} also applies to realistic superconducting qubit setups with additional inductive couplings between nearest neighbors~\cite{zanner2021coherent,Ustinov2021}. 

{A related decomposition  exists also for the non-Hermitian effective Hamiltonian $H_{L}=-\rmi c_L^\dag c_L^{\vphantom{\dag}}$,
namely $H_{L}=-\rmi (\widetilde{\mathcal{F}}+\widetilde{\mathcal B}^T \widetilde{\mathcal B}) $. This proves that the imaginary parts of all the eigenvalues of $H_L$ in the subspace of words $w$ with a given  $s_w$ are below $-|J_x|/2$ where $J_x=s_w - (N/2)$. The explicit form of the diagonal matrix 
$\widetilde{\mathcal{F}}$ and the incidence matrix $\widetilde{\mathcal B}$, describing the Laplacian of simplicial complex $\widetilde{\mathcal B}^T\widetilde{\mathcal B}$,  is given in the End Matter. It is the study of this simplified problem for $H_L$ that has inspired our decomposition of the full Liouvillian superoperator.}

{\it Summary and outlook.}
To summarize, we demonstrate that subradiance is suppressed by a strong coherent driving by proving a general spectral bound for the decay rate of the correlations in an atomic array coupled to photons in a waveguide. Our results are based on the exact decomposition of the Liouvillian superoperator into a diagonal operator and a  Laplacian of a ranked poset.
The proposed technique is rather general and can hopefully be applied to other atom-photon setups beyond waveguide QED. It could be instructive to look for oscillating subradiant correlations in more complex systems, for example, beyond one spatial dimension.

\begin{acknowledgements}
We thank Shira Barnoy-Lapid, Oren Raz and Xin Zhang for useful discussions. 
The work of
JS and ANP has been supported by the Simons Foundation, by the research grants from the
Center for New Scientists and from the Center for Scientific
Excellence at the Weizmann Institute of Science, and by the
Quantum Science and Technology Program of the Israel
Council for Higher Education. RT was supported by the ISF grant No.~1729/23.
\end{acknowledgements}
%



 \setcounter{equation}{0}
 \renewcommand{\theequation}{A\arabic{equation}}


 \onecolumngrid
 \newpage\clearpage 
 \begin{center}{\bf End Matter}\end{center}
 \twocolumngrid
\appendix
\section{The Effective Hamiltonian}

In this section, we study the limits on eigenvalues of an effective Hamiltonian $H_{L}=-\rmi c_L^\dag c_L^{\vphantom{\dag}}$ in the limit of strong driving. In order to introduce the driving, we consider the effect of the Hamiltonian in the subspaces spanned by  the eigenstates $|w\rangle$ of $\hat{J_x}$ with the same eigenvalue $J_x= s_w - (N/2)$. Without the loss of generality, we consider only the cases $N/2 \leq s_w < N$. In the reduced basis $\{ |w\rangle\}$, $c_L^\dag c_L^{\vphantom{\dag}}$ are represented by a matrix $\widetilde{\mathcal{Q}} = \langle w' |c_L^\dag c_L^{\vphantom{\dag}}|w\rangle$:
\begin{align}\label{eq:matrixRR}
    \widetilde{\mathcal{Q}}_{w w'} = \frac{1}{2} &\begin{cases}
\sum\limits_{j\ge k}(-1)^{w_j+w_k}\cos(\phi_j-\phi_k), & (w=w')\\
\cos(\phi_j-\phi_k), & (w\:\&\: w' \text{ differ } 
\end{cases}\nonumber\\
&\hspace{3.2cm}\text{only in  indices }j\text{ and }k)\:.
\end{align}
We now decompose this matrix as a sum of a diagonal matrix $\widetilde{\mathcal{F}}$ and a Laplacian $\widetilde{\mathcal{B}}^T \widetilde{\mathcal{B}}$, $\widetilde{\mathcal{Q}} = \widetilde{\mathcal{F}} + \widetilde{\mathcal{B}}^T \widetilde{\mathcal{B}}$. {Formally, $\widetilde{\mathcal{B}}^T \widetilde{\mathcal{B}}$ is a Laplacian of a simplicial complex $\{w\}$}.
It is straightforward to notice that the elements of $\widetilde{\mathcal{B}}$ are determined by the words ${w}$ and ${v}$ with {$r(v,w)=1$}. If $v_i$ differs from  $w_j$ by exactly one index $k$, then $\widetilde{\mathcal{B}}_{ij} = \boldsymbol{\theta}_k$. Otherwise $\widetilde{\mathcal{B}}_{ij} = 0$. 

If the words $w$ and $w'$ differ only in the indices $j$ and $k$ with $w_j=0$ and $w'_k=0$, there is only one word $v$ that differs from $w$ in index $j$ and from $w'$ in index $k$. The elements of matrix $\widetilde{\mathcal{B}}$ then are $\widetilde{\mathcal{B}}_{v,w} = \boldsymbol{\theta}_j$ and $\widetilde{\mathcal{B}}_{v,w'} = \boldsymbol{\theta}_k$ with the remaining elements on those two columns being zero. Finally, $\widetilde{\mathcal{B}}^T \widetilde{\mathcal{B}}$ results in $\widetilde{\mathcal{Q}}_{w w'} = \boldsymbol{\theta}_j^T \cdot \boldsymbol{\theta}_k =  \cos(\phi_j-\phi_k)/2$.
The diagonal terms of $\widetilde{\mathcal{B}}^T \widetilde{\mathcal{B}}$ are $(N-s_w)/2$, which makes 
\begin{equation}
    \widetilde{\mathcal{F}}_{w} = \frac{|a_w|^2}{4} + \frac{2s_w-N}{4}\:,
\end{equation}
where $a_w = \sum_{j=1}^N (-1)^{w_j} \e^{\rmi \phi_j}$.
It can be shown that for generic values of $\varphi_j$ the matrix $\widetilde{\mathcal{Q}}$ has a one dimensional kernel, spanned by the eigenvector
\begin{equation}\label{eq:psi0}
[\psi_{0}]_{w}=
\prod\limits_{j>k,w_{j}=w_{k}}
\sin(\phi_{j}-\phi_{k})(-1)^{w_{j}+j+k}\:.
\end{equation}
This eigenvector transforms according to the sign representation of the permutation group built on the vector space spanned by eigenstates $|w\rangle$ and coefficients in the ring of polynomials in $\sin(\varphi_i),\cos(\varphi_i),$ where the symmetric group $S_N$ acts by simultaneously permuting the indices $1,2,\ldots, N$ and the variables $\phi_1,\ldots,\phi_N$.

\vspace{0.5cm}
\section{A proof of the equality $\mathcal{A}=\mathcal{B}^T\mathcal{B}$}

We now present an explicit proof that the matrix $\mathcal{A}$ in the main text is equal to $\mathcal{B}^T \mathcal{B}$. We notice that
\begin{equation}
    [\mathcal{B}^T \mathcal{B}]_{ij} = \sum_k \mathcal{B}_{ki} \mathcal{B}_{kj}\:,
\end{equation}
where $i$ and $j$ correspond to $i$-th and $j$-th words in $\{(w,w')\}$ while $k$ enumerates words in $\{(t,t')\}$,   ${r(w,w') =m}$ and ${r(t,t') =m+1}$. If $i=j$, the diagonal terms are determined by the number of nonzero elements $\mathcal{B}_{ki}$ since $\boldsymbol{\theta}_j^T \cdot \boldsymbol{\theta}_k =1/2$. Given a word pair $(w,w')$, one can construct the required $(t,t')$ by flipping a $0$ to a $1$ in the left word or flipping a $1$ to a $0$ in the right word. There are $N-m$ number of possibilities, so the diagonal terms of $\mathcal{B}^T \mathcal{B}$ are $(N-m)/2$.

As for the off-diagonal terms, we are looking for the word pair $(t,t')$ that makes both $\mathcal{B}_{ki}$ and $\mathcal{B}_{kj}$ nonzero. By construction of the matrix $\mathcal{B}$, $(w,w')_i$ and $(w,w')_j$ must be obtained from the same $(t,t')$ by flipping a $1$ to $0$ in the left words or flipping a $0$ to $1$ in the right word. We now define $(p,p') \equiv (w,w')_i$ and $(q,q')\equiv (w,w')_j$.

If $p'=q'=t'$, the problem is reduced to the off-diagonal terms of the matrix $\widetilde{\mathcal{Q}}$ in Eq.~\eqref{eq:matrixRR}. When $p$ and $q$ differ only in indices $x$ and $y$,  $[\mathcal{B}^T \mathcal{B}]_{ij} = \frac{1}{2} \cos(\phi_x-\phi_y)$. Otherwise $[\mathcal{B}^T \mathcal{B}]_{ij} = 0$. The same argument holds for $p=q=t$. They match the second and third terms in Eq.~\eqref{eq:matrixA}.

The last possible situation is that one word pair, assumed to be $p$, is obtained by flipping a $1$ to $0$ on index $x$ of the word $t$, and the other word pair $q$ is obtained by flipping a $0$ to $1$ on index $y$ of the word $t'$. Therefore, $\mathcal{B}_{ki} = \boldsymbol{\theta}_x$ and $\mathcal{B}_{kj} = -\boldsymbol{\theta}_y$, where $k$ is the index of $(t,t')$ in set $\{(t,t')\}$. Finally, $[\mathcal{B}^T \mathcal{B}]_{ij} = -\frac{1}{2} \cos(\phi_x-\phi_y)$. This matches the first term in Eq.~\eqref{eq:matrixA}.

This concludes the proof of the identity $\mathcal{A} = \mathcal{B}^T \mathcal{B} $.


\end{document}